\documentclass[aps,prl,twocolumn,superscriptaddress,showpacs,footnotebib]{revtex4-1}
\usepackage{amsmath}
\usepackage{bm}
\usepackage{color}
\usepackage{graphicx}
\usepackage{epstopdf}
\usepackage{epsfig}
\usepackage{amsfonts}
\usepackage[naturalnames]{hyperref}
\usepackage{hypcap}
\usepackage{verbatim}
\usepackage{tabularx}
\usepackage{bbm}
\usepackage{esvect}
\usepackage{subfigure}

\def\bea{\begin{eqnarray}}
\def\eea{\end{eqnarray}}
\def\nn{\nonumber}
\def\ba{\begin{array}}
\def\ea{\end{array}}
\def\Tr{\text{Tr}}
\def\nn{\nonumber}

\hypersetup{colorlinks=true, citecolor=blue, urlcolor=blue, linkcolor=blue}
\begin{document}

\title{Charge-$4e$ Superconductivity from Nematic Superconductors in 2D and 3D}

\author{Shao-Kai Jian}
\affiliation{Condensed Matter Theory Center, Department of Physics, University of Maryland, College Park, Maryland 20742, USA}

\author{Yingyi Huang}
\affiliation{Institute for Advanced Study, Tsinghua University, Beijing 100084, China}
\affiliation{School of Physics and Optoelectronic Engineering, Guangdong University of Technology, Guangzhou 510006, China}

\author{Hong Yao}
\email{yaohong@tsinghua.edu.cn}
\affiliation{Institute for Advanced Study, Tsinghua University, Beijing 100084, China}
\affiliation{State Key Laboratory of Low Dimensional Quantum Physics, Tsinghua University, Beijing 100084, China}

\date{\today}

\begin{abstract}
    Charge-$4e$ superconductivity as a novel phase of matter remains elusive so far.
    Here we show that charge-4e phase can arise as a vestigial order above the nematic superconducting transition temperature in time-reversal-invariant nematic superconductors. 
    On the one hand, the nontrivial topological defect---nematic vortex---is energetically favored over the superconducting phase vortex when the nematic stiffness is less than the superfluid stiffness; consequently the charge-$4e$ phase emerges by proliferation of nematic vortices upon increasing temperatures. 
    On the other hand, the Ginzburg-Landau theory of the nematic superconductors has two distinct decoupling channels to either charge-$4e$ orders or nematic orders; 
    by analyzing the competition between the effective mass of the charge-$4e$ order and the cubic potential of the nematic order, we find a sizable regime where the charge-$4e$ order is favored.
    These two analyses consistently show that nematic superconductors can provide a promising route to realize charge-$4e$ phases, which may apply to candidate nematic superconductors such as PbTaSe$_2$ and twisted bilayer graphene.
\end{abstract}
\maketitle

{\it Introduction.---}Featuring the condensation of quartets with four times the fundamental electron charges, charge-$4e$ superconductivity~\cite{korshunov1985two, kivelson1990doped,wu2005competing,berg2009charge,jiang2017charge} is intrinsically distinct from the conventional charge-$2e$ superconductivity~\footnote{In this paper, we refer the charge-$2e$ superconductivity as conventional superconductivity to distinguish it from the charge-$4e$ superconductivity. For example, both $s$-wave isotropic superconductor and $d$-wave nematic superconductor are conventional in this context.} discovered more than a century ago. 
One hallmark of charge-4e superconductors is the magnetic flux quantization with period $hc/4e$, which is half that of the usual superconducting flux quantum. 
Unlike the conventional superconductivity that is well described by the seminal Bardeen–Cooper–Schrieffer (BCS) theory, many properties of charge-$4e$ superconductivity remain less well understood.
The mystery of this phase is not only because it defies any BCS-type analysis, but also due to the lack of experimental realization so far. 
Previous studies in Ref.~\cite{berg2009charge} suggested that the charge-$4e$ superconductivity can arise as a vestigial long-range order above the transition temperature of a pair-density-wave (PDW) superconductor~\cite{fulde1964superconductivity, larkin1965zh,radzihovsky2009quantum, Agterberg2008NPdislocations, wang2010spin, cho2012superconductivity, lee2014amperean, maciejko2014weyl,jian2015emergent,jian2017emergence,jian2020mass,Han2020PRL-PDW,kivelson2020-arcmp}, whose order parameter varies periodically in the real space.
The additional breaking of translational symmetry in the PDW state is essential in realizing the charge-$4e$ superconductivity as it provides nontrivial topological defects---dislocations. 
As a consequence, if the dislocation proliferates as the temperature is raised, the equilibrium state will restore the translational symmetry by doubling the original charge-$2e$ condensate, and therefore lead to the charge-$4e$ phase.

The crucial ingredient of the underlying condensate having more than one quantum number~\cite{chung2012charge, Moon2012PRB-charge4e, xu2018topological, takeuchi2020phase} provides a general guide to search for charge-$4e$ superconductors.
It was proposed that in certain two-dimensional high-temperature superconductors (e.g. La$_{2-x}$Ba$_x$CuO$_4$ and La$_{1.6-x}$Nd$_{0.4}$Sr$_x$CuO$_4$~\cite{fujita2004stripe, tranquada2008evidence, hucker2011stripe}) exhibiting stripe superconducting orders~\cite{berg2007dynamical, berg2009theory}, the charge-$4e$ superconductivity may occur above the transition temperature~\cite{berg2009charge}.
Nevertheless, the experimental signature of the charge-$4e$ superconductivity in these high-temperature superconductors has not been observed so far.
Apart from PDW superconductors, the nematic superconductor that breaks both charge conservation and lattice rotational symmetry also hosts an order parameter that carries multiple quantum numbers.
The additional topological defect in nematic superconductors is created by the intersection of domain walls separating different ground states.  
More importantly, many experimental progresses have been made in achieving the nematic superconductivity in various systems.
For instance, recent experiments found that the spin susceptibility below the superconducting temperature breaks the three-fold lattice rotational symmetry in doped topological insulator Cu$_x$Bi$_2$Se$_3$~\cite{matano2016spin,yonezawa2017thermodynamic,tao2018direct, yonezawa2019nematic}, where the breaking of rotational symmetry suggests that the order parameter in the superconducting phase is a two-dimensional $E_u$ representation of point group $D_{3d}$~\cite{fu2014odd,venderbos2016odd, hecker2018vestigial}. 
Providing the growing experimental evidence of the nematic superconductivity in various systems, including doped topological insulators Cu$_x$Bi$_2$Se$_3$~\cite{matano2016spin, yonezawa2017thermodynamic,tao2018direct, yonezawa2019nematic}, Sr$_x$Bi$_2$Se$_3$~\cite{wang2019evidence, kostylev2020uniaxial} and Nb$_x$Bi$_2$Se$_3$~\cite{shen2017nematic, asaba2017rotational}, superconducting topological semimetals PbTaSe$_2$~\cite{XinLu2020evidence}, and more recently twisted bilayer graphene~\cite{kerelsky2019maximized,  cao2020nematicity}, as well as increasing interest in novel charge-$4e$ ordering, it is thus interesting to ask whether the multiple-component superconducting order parameters in nematic superconductors is able to realize charge-$4e$ phases above their charge-$2e$ transition temperature.

\begin{figure}
\centering
\includegraphics[width=5cm]{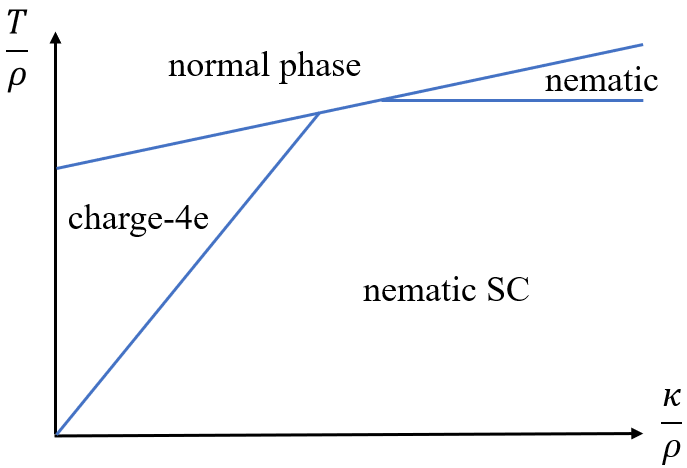}
\caption{Schematic phase diagram of the defect theory. $\kappa$, $\rho$ and $T$ denote the nematic stiffness, the superfluid stiffness and temperature, respectively. The solid lines refer to phase boundaries. 
The three-fold anisotropy is relevant at the transition point from the charge-$4e$ superconductor to the nematic superconductor.
\label{fig:RG}}
\end{figure}

In this paper, we answer this question in the affirmative; namely, we provide two analyses, suitable for 2D and 3D, respectively, to support the possibility of a vestigial charge-$4e$ superconducting phase from nematic superconductors. 
In 2D, because a vortex is a point-like object, we analyze the fate of various topological defects to determine the phase diagram. 
After identifying three distinct topological defects that are responsible for three different orders out of the nematic superconducting phase, 
we find that the competition between the superfluid stiffness and the elastic constant leads to a rich phase diagram, as shown in Fig.~\ref{fig:RG}. 
In particular, if the elastic constant of the material is less than one third of the superfluid stiffness, the energetically favored nematic vortices are proliferated, resulting in the novel charge-$4e$ phase, when the temperature gets raised above the transition temperature.
In 3D, we employ a Ginzburg-Landau theory near the transition point where the nematic superconducting order parameter is well described by a simple field theory up to quartic terms.
We are able to show analytically that the effective mass of the charge-$4e$ order is less than that of the nematic order by treating the three-fold anisotropy perturbatively, which indicates the charge-$4e$ order is favored. 
We further confirm our results by numerically solving the saddle-point equation.
Our result in 3D suggest that materials whose dispersion along the third direction is weaker than the in-plane dispersion such as in Sr$_x$Bi$_2$Se$_3$ are promising in realizing the charge-$4e$ phase. 

{\it Charge-4$e$ phase from proliferating topological defect.---}We start by analyzing the symmetries in the nematic superconducting phase providing an appropriate language for clarifying the topological defects, and then turn to renormalization group analysis of the defect theory.
As mentioned above, we consider the order parameter of the nematic superconducting phase being a two-component complex boson $\Delta= (\Delta_x, \Delta_y)^T$ carrying $E_u$ representation in $D_{3d}$ group, and each complex component additionally hosts the $U(1)$ quantum number corresponding to the charge conservation, i.e., it is the conventional charge-$2e$ condensate. 
In the phase respecting time-reversal symmetry, the relative phase between two components $\Delta_x, \Delta_y$ is pined at $0$ or $\pi$. 
As we focus on the time-reversal-invariant nematic superconductor, we hereafter assume the relative phase between $\Delta_x$ and $\Delta_y$ is pinned at $0$ in the analysis of low-energy physics.

In terms of the basis $\Delta_\pm = \Delta_x \pm i \Delta_y$, one can bring the the four-dimensional field configuration into three phase modes and one amplitude mode.
The superconducting phase difference between two components represents the degree of time reversal symmetry breaking. 
Because the nematic superconducting state respects the time reversal symmetry, the phase difference mode can be ignored in the low energy limit; namely, its fluctuation is always gapped in our analysis.  
Thus, we have two phase modes and one amplitude mode,
\bea \label{eq:order}
	\Delta_+= |\Delta| e^{ i(\theta+ \phi)} , \quad \Delta_-= |\Delta| e^{ i(\theta-\phi)}.
\eea
where $\theta$ and $\phi$ denote the two phase modes.
The $\phi$ field describes the $U(1)$ rotation between amplitudes of two components (i.e. the spatial rotation), whereas the $\theta$ field is the $U(1)$ phase conjugated to the global charge.
The two $U(1)$ phases can also be understood by secondary orders, i.e., the nematic order $Q \sim \Delta_-^\dag \Delta_+$ and the charge-$4e$ order $\Delta_{4e} \sim \Delta_+ \Delta_-$, that transform as
\bea \label{eq:transform}
	 Q  \rightarrow  Q e^{2i\phi}, \quad \Delta_{4e} \rightarrow \Delta_{4e} e^{2i\theta}. 
\eea

The single-valueness of nematic superconducting order parameter $\Delta_\pm$ uniquely determines that the topological defects are given by $(\delta\theta, \delta\phi)= (2\pi,0), (0,2\pi), (\pi,\pi)$, where $\delta \theta$, $\delta \phi$ denote the winding of the phase around a defect. 
Physically, they correspond to the superconducting vortex, the nematic double vortex and the superconducting half-vortex binding with a single nematic vortex, respectively. 
Deep in the nematic superconducting phase, the effective action characterizing the phase modes is~\cite{berg2009charge}
\bea
	S = \int d^2 r \left( \frac\rho{2T} (\partial \theta)^2+ \frac\kappa{2T} (\partial \phi)^2 - g_6 \cos 3 \phi \right),
\eea
where $\rho$ is the superfluid stiffness (superfluid density), $\kappa$ is the nematic stiffness (elastic constant), and $T$ denotes the temperature.
Note that $\cos 3 \phi$ is allowed owing to the three-fold anisotropy, i.e., the action is invariant under $\phi \!\rightarrow\! \phi \!+\!2\pi/3$. 
$g_6$ is the strength characterizing the three-fold anisotropy.
Thermal proliferation of topological defects can be then described by the dual bosons $\tilde \phi$ and $\tilde \theta$~\cite{berg2009charge},
\bea 
	S &=& \int d^2 x \Big( \frac{T}{2\rho} (\partial \tilde\theta)^2+ \frac{T}{2\kappa} (\partial \tilde\phi)^2-  g_6 \cos 3 \phi - g_{2,0} \cos2\pi\tilde\theta \nn \\
	&& - g_{0,2} \cos2\pi\tilde\phi- g_{1,1} \cos\pi\tilde\theta \cos\pi\tilde\phi, \Big) \label{eq:defect_H}
\eea
where $g_{2,0}$, $g_{0,2}$ and $g_{1,1}$ are couplings characterizing the strength of creating (annihilating) each kind of topological defect, respectively.

The standard renormalization group flow for Eq.~(\ref{eq:defect_H}) to the lowest order reads
\bea
	\frac{d g_{2,0}}{dl} &=& \left(2- \frac{\pi \rho}T \right) g_{2,0}, \\
	\frac{d g_{0,2}}{dl} &=& \left(2- \frac{\pi \kappa}T \right) g_{0,2}, \\
	\frac{d g_{1,1}}{dl} &=& \left[2- \frac{\pi }{4T}(\rho+ \kappa) \right] g_{1,1}, \\
	\frac{d g_6}{dl} &=& \left(2- \frac{9 T}{4\pi \kappa} \right)g_6.
\eea
The coefficient in front of the coupling on the right-hand side of the renormalization group equation determines whether the corresponding process is relevant or not.
For instance, if $\frac{T}{\rho} > \frac{\pi}2$, the creation and annihilation process of the superconducting vortex is relevant, leading to the proliferation of superconducting phase defects. 
Consequently, the condensation of the dual field destroys the superconducting phase coherence, driving the system out of superconducting orders.
The positions where the creation and annihilation operators become marginal determines the phase boundary. 
In this way, the phase diagram as shown in Fig.~\ref{fig:RG} is mapped out by the renormalization group equations. 
It is worth noting that the proliferation of nematic vortices at $\frac{T}\kappa = \frac2{\pi}$ restores the lattice rotational symmetry though the anisotropy term $g_6$ seems relevant above the transition point $\frac{T}\kappa > \frac2{\pi}$. This is an artifact of the one-loop calculation: the anisotropy term is actually irrelevant above the transition point~\cite{wu1982potts}.
As a result the charge-$4e$ transition is given by $\frac{T}\kappa = \frac2{\pi}$
when the nematic stiffness is less than the superfluid stiffness, more specifically $\kappa < \frac{\rho}{3} $. 
Physically, this is because it is easier to create nematic vortices than superconducting vortices such that raising temperature can more efficiently proliferate the nematic vortices, restoring the lattice rotational symmetry but not the $U(1)$ charge symmetry. 
It realizes a charge-$4e$ order because the charge-$4e$ order is 
not visible to and then not directly affected by the nematic vortex (i.e., it respects the lattice rotational symmetry), as shown by the symmetry transformation law in Eq.~(\ref{eq:transform}), and because any charge-$2e$ nematic superconducting order breaks lattice rotational symmetry and is thus melted by the proliferation of nematic vortices. 

Besides the interesting charge-$4e$ phase, the competition between the nematic stiffness and the superfluid stiffness results in a rich phase diagram in Fig.~\ref{fig:RG}.
Namely, when $1/3 < \frac\kappa\rho < 3$, raising the temperature causes a direct transition from the nematic superconductivity to the normal phase since proliferating a superconducting half-vortex bounded with a nematic vortex is favored, whereas, when $ \frac\kappa\rho > 3$, a vestigial of nematic phase emerges since proliferating the normal superconducting phase vortex is favored~\cite{cho2020z}.

{\it Charge-$4e$ phase from superconducting fluctuations.---} In 3D, the Ginzburg-Landau theory works better since the quantum fluctuation is generally suppressed as dimension increases, so we expect a mean-field analysis to 3D nematic superconductivity near the phase boundary can describe the essential physics.
Such a theory was also used in Ref.~\cite{hecker2018vestigial} to investigate the vestigial nematic order, but the authors did not analyze the possibility of charge-$4e$ orders. 
The Ginzburg-Landau theory of the nematic superconducting order parameter $\Delta= (\Delta_x, \Delta_y)^T$ near the phase boundary reads~\cite{hecker2018vestigial}
\bea \label{eq:nematic_SC}
	S = \int_p \Delta^\dag G_p^{-1}  \Delta+ \int_x( u (\Delta^\dag \Delta)^2 + v (\Delta^\dag \tau^y \Delta)^2),
\eea
where $\int_p = \int \frac{dp^3}{(2\pi)^3}$, $\int_x = \int d^3 x$, and
\bea \label{eq:inverse_propagator} 
G_p^{-1} = m_0(\bm{p})\tau^0  +m_1(\bm{p})\tau^z +m_2(\bm{p})\tau^x,
\eea
is the inverse propagator of the nematic superconducting order parameter with 
$m_0(\bm{p})=d_\parallel (p_x^2 + p_y^2) + d_z p_z^2 + r_0$, $m_1(\bm{p})=d' (p_x^2 - p_y^2)+ \bar d p_y p_z$, and $m_2(\bm{p})=d' 2p_x p_y+ \bar d p_x p_z$.
Notice that while $m_0(\bm p)$ enjoys a continuous rotational symmetry, $m_1(\bm p)$ and $m_2(\bm p)$ lower it down to three-fold rotation.
$r_0$, $u$, $v$ are real parameters allowed by the symmetry and $d_\parallel$, $d_z$, $d'$, $\bar d$ characterize the kinetic energy.
In the following, we consider $v>0$, $u>0$, as the case for $v<0$ favors the chiral superconductivity that breaks time reversal symmetry. 
It is crucial to realize that there are two ways given by the following two different Fierz identities to decouple the last quartic term in Eq.~(\ref{eq:nematic_SC})
\bea
\label{eq:fierz_nematic}    \tau^y_{\alpha\beta} \tau^y_{\gamma\delta} &=& 2 \delta_{\alpha\delta} \delta_{\beta\gamma} - \delta_{\alpha\beta} \delta_{\gamma\delta}- \tau^x_{\alpha\beta} \tau^x_{\gamma\delta} - \tau^z_{\alpha\beta} \tau^z_{\gamma\delta}, \\
\label{eq:fierz_4e}    \tau^y_{\alpha\beta} \tau^y_{\gamma\delta} &=& \delta_{\alpha\beta} \delta_{\gamma\delta} - \delta_{\alpha \gamma} \delta_{\beta\delta}.
\eea
This leads to either nematic channel for the first one Eq.~(\ref{eq:fierz_nematic}) or charge-$4e$ channel for the second one Eq.~(\ref{eq:fierz_4e}). 
After the decoupling, one is able to integrate out the quadratic nematic superconducting order parameter.
We leave the details to the Supplemental Material~\cite{SM}, and present the main results here.
The Ginzburg-Landau theory for the nematic order and the charge-$4e$ order (assuming to be homogeneous in real space) are given by
\bea
\label{eq:nematic_action}     S_\text{nem} &=& \Tr\log \chi_p^{-1} + \int_x \left( \frac1{4v} (Q_x^2 + Q_y^2) - \frac1{4u'} R^2 \right),~~~ \\
\label{eq:4e_action}     S_{4e} &=& \frac12 \Tr \log D_p^{-1}  + \int_x \left( \frac1{4v} |\Delta_{4e}|^2 - \frac1{4u'} R^2 \right),
\eea
where $u'= u +v$, $Q_x, Q_y$ are the nematic orders, $\Delta_{4e}$ is the charge-$4e$ order, and $R \sim \Delta^\dag \Delta$ is a scalar that decouples the first quartic interaction in Eq. \eqref{eq:nematic_SC}. 
The $\Tr \log $ term comes from integrating out the nematic superconducting order parameter, and
$\chi_p^{-1} = G_p^{-1} + R + Q_x \tau^z + Q_y \tau^x$, and $D_p^{-1} =
G^{-1}_p + R +  \Delta_{4e} \rho^+ +  \Delta_{4e}^\ast \rho^-$ are the modified inverse propagators in the presence of the nematic order and the charge-$4e$ order, respectively. 
$\rho^{\pm} = \frac12( \rho^x \pm i \rho^y)$ where $\rho^x$ and $\rho^y$ are Pauli matrices acting on the ``Nambu" space, i.e., $\Delta_{4e}$ and $\Delta_{4e}^\ast$, and the extra factor $\frac12$ in front of the $\Tr \log$ term in Eq.~(\ref{eq:4e_action}) is due to the redundancy of the enlarged Nambu space. 

The Fierz identity and the bosonic nature of the nematic superconducting order parameter $\Delta$ cause the similar Ginzburg-Landau theory for the nematic order Eq.~(\ref{eq:nematic_action}) and the charge-$4e$ orders Eq.~(\ref{eq:4e_action}).
The only difference comes from the $\Tr \log $ term as the nematic order carries a real space (nematic) quantum number while the charge-$4e$ order carries a Nambu space quantum number.
Unlike the fermionic field, the bosonic nematic superconducting field is blind to the Nambu space as one can easily observe that the inverse propagator $G_p^{-1}$~(\ref{eq:inverse_propagator}) is an identity in the Nambu space. 
Thus if one artificially turns off $m_1$ and $m_2$ such that the nematic superconducting field has an accidental continuous real space rotational symmetry at the leading order, the nematic order and the charge-$4e$ order will be degenerate since now the nematic superconducting field is blind to both nematic space and Nambu space (the factor of $\frac12$ in Eq.~(\ref{eq:4e_action}) is perfectly compensated by the enlarged Nambu space).
So even without any calculation, one can identify a specific albeit artificial point where the charge-$4e$ order is degenerate with the nematic order.
Now the question is how the nonvanishing $m_1$ and $m_2$, equivalently $d'$ and $\bar d$, affect the two instabilities. 
In the following, we provide an analytical result showing the charge-$4e$ order is favored over the nematic order when $d' \gg \bar d$, which makes quasi-2D materials whose out-of-plane dispersion is much weaker than the in-plane dispersion promising candidates for realizing charge-$4e$ phases.
In addition, we also numerically solve the saddle-point equation for both nematic orders and charge-$4e$ orders, which confirms the analytical result.

\begin{figure}
\subfigure[]{\label{fig:feyn_4e}
    \includegraphics[width=2.4cm]{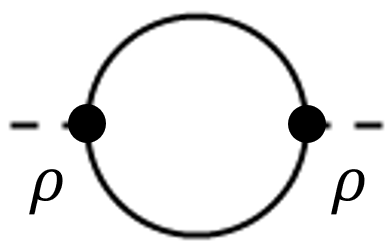}} \qquad \qquad
\subfigure[]{\label{fig:feyn_nematic}
    \includegraphics[width=2.4cm]{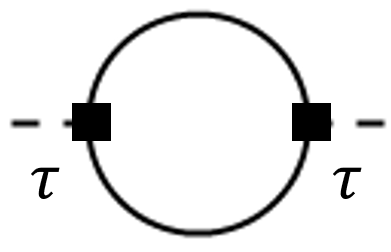}}
    \caption{The Feynman diagrams that contribute to the effective mass of the charge-$4e$ order and the nematic order, respectively. The difference of (a) and (b) arises from the different vertices denoted by $\bm\rho$ and $\bm\tau$. }
    \label{fig:feyn}
\end{figure}

Near the transition point, the effective Ginzburg-Landau theory can be obtained by assuming order parameters small and expanding the action order by order. 
The effective mass that is crucial in determining the phase boundary is given by evaluating the Feynman diagram in Fig.~\ref{fig:feyn}, and as we discussed above the difference lies in the vertices of the two orders as shown in Fig.~\ref{fig:feyn_4e} and Fig.~\ref{fig:feyn_nematic}.  
Since nonvanishing $m_1$ and $m_2$ are diagonal in the Nambu space the contributions from the $\tau^z$ and $\tau^x$ kinetic energy are additive to the charge-$4e$ order's mass. 
Note also, the correction is negative in general since it is a boson bubble. 
This means there is a tendency to form the charge-$4e$ order above the transition temperature of the nematic superconductivity. 
On the other hand, since nonvanishing $m_1$ and $m_2$ are not diagonal in the nematic space the contributions tend to cancel each other. 
We calculate the effective mass perturbatively in $\bar d$, and the lowest-order result is
\bea
    m_{4e} &=& m_\text{nem}  - \delta m_2, \\
    \delta m_2 &=& \frac1{ d_\parallel \sqrt{d_z (R+r_0)}} \frac{2\gamma-(1- \gamma^2)\log\left( \frac{1+\gamma}{1-\gamma} \right)}{32\pi \gamma(1-\gamma^2)},
\eea
where $m_{4e}$ and $m_\text{nem}$ denote the effective mass of the charge-$4e$ order and the nematic order, respectively, and $\gamma = \frac{d'}{d_\parallel}$. 
It is not hard to see that $\delta m_2$ is positive and monotonic in $0< \gamma <1$, indicating the anisotropy induced by nonvanishing $d'$ favors the charge-$4e$ order.

However, this enhancement of the charge-$4e$ instability needs to compete with the cubic potential of nematic orders arisen from the three-fold anisotropy. 
More concretely, a cubic term of the nematic order, i.e., $S_\text{nem}^{(3)} = w \int_x ( Q_x^3 - 3 Q_x Q_y^2 )$, is present in the expansion of Eq.~(\ref{eq:nematic_action}). 
The cubic term appears at least at the quadratic order in $\bar d$ and the linear order in $d'$ (actually, the dependence of $d'$ at quadratic order of $\bar d$ can be obtained explicitly), and through explicit calculation, the cubic term is 
\bea \label{eq:cubic}
    S_\text{nem}^{(3)} &=& w \int_x ( Q_x^3 - 3 Q_x Q_y^2 ), \\
    w &\approx& \frac1{960 \pi} \frac1{[d_z (R+r_0)]^{3/2}} \frac{\bar d^2 d'}{d_\parallel^3}.
\eea
It is a simple matter of fact that the cubic term enhances the instability at least in the quadratic orders, i.e., $\delta m_3 \propto w^2 \propto \bar d^4$, so the lowest order is the quartic order in $\bar d$.
On the other hand, the enhancement of the charge-$4e$ superconductivity $\delta m_2$ appears already in the zeroth order of $\bar d$.
For the systems described by Eq.~(\ref{eq:nematic_SC}), there exists a parameter regime, where $\bar d \ll d'$, such that the charge-$4e$ order is favored over the nematic order.
Notice that we have neglected the sixth order terms like $ \left( \Delta^\dag (\tau^z + i \tau^x) \Delta \right)^3 + h.c. $ in Eq.~(\ref{eq:nematic_SC}) that are allowed by the three-fold anisotropy, and these terms can also render cubic contributions to the nematic order effective action Eq.~(\ref{eq:cubic}). 
These higher order terms are in general irrelevant for the long wavelength physics near and slightly above the nematic superconducting transition temperature at which the Ginzburg-Landau theory Eq.~(\ref{eq:nematic_SC}) can apply.
(Whereas if the temperature is lower than the transition temperature, this becomes dangerously irrelevant~\cite{jian2020mass} and selects one of the degenerate ground states). 
Unlike the universal physics that can be obtained with the help of the Ginzburg-Landau action in Eq.~(\ref{eq:nematic_SC}), it is not clear away from the applicability of such an action what the fate of the competition between the charge-$4e$ order and the nematic order is. 
Nevertheless, the 2D calculation shown in Fig.~\ref{fig:RG} provides a complementary qualitative understanding: the large anisotropy from higher-order potentials can increase the nematic stiffness and prohibit the charge-$4e$ phase~\cite{cho2020z}.

\begin{figure}[t]
    \subfigure[]{    
        \includegraphics[width=4cm]{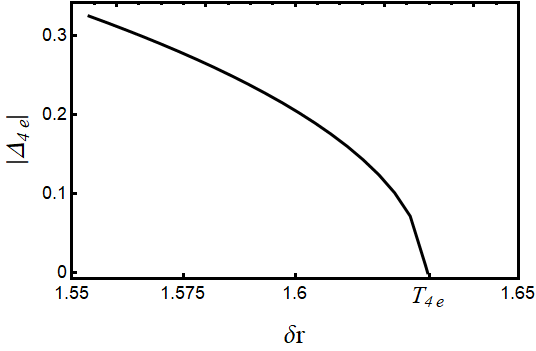}} \quad
    \subfigure[]{
        \includegraphics[width=4cm]{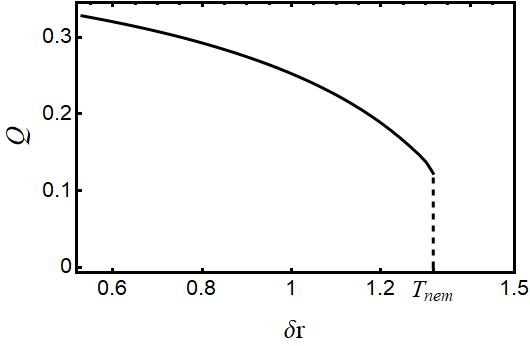}}
    \caption{The mean field phase diagram of the charge-$4e$ order (a) and the nematic order (b). $|\Delta_{4e}|$, and $Q= \sqrt{Q_x^2 + Q_y^2}$ are the amplitude of the charge-$4e$ and the nematic order respectively. $\delta r = r_0 - r_0^\ast \propto T-T^\ast$, where $r^\ast$ ($T^\ast$) denotes the transition point (transition temperature) without vestigial orders. The parameters are $d_\parallel=1, d'=0.5, d_z = 0.1, \bar d=0.1, v=1, u=4$. }
\label{fig:meanfield}
\end{figure}

To further support our result, we carry out a numerical calculation of the saddle-point equations in both the charge-$4e$ channel and the nematic channel.
We leave the saddle-point equation in the Supplemental Material~\cite{SM}. 
The phase diagram is shown in Fig.~\ref{fig:meanfield}, where $|\Delta_{4e}|$, and $Q= \sqrt{Q_x^2 + Q_y^2}$ are the amplitude of the charge-$4e$ and the nematic order respectively. $\delta r = r_0 - r_0^\ast \propto T-T^\ast$, where $r^\ast$ ($T^\ast$) denotes the transition point (transition temperature) without vestigial orders.
That the system is in favor of the charge-$4e$ phases when $\bar d \ll d'$ is demonstrated explicitly by $T_\text{nem}<T_{4e}$ in Fig.~\ref{fig:meanfield}.
One should also notice the first-order jump of the nematic transition due to the cubic anisotropy, on the contrary, the phase transition to the nematic phase at two dimensions is continuous due to the strong quantum fluctuations~\cite{wu1982potts}.

{\it Conclusions.---}We have revealed the promising possibility of realizing a novel charge-$4e$ superconducting phase above the transition temperature of time-reversal-invariant nematic superconductors in 2D and quasi-2D. 
Since experimental evidence of rotational symmetry breaking superconducting phases in various systems has accumulated, including doped topological insulators Cu$_x$Bi$_2$Se$_3$~\cite{matano2016spin, yonezawa2017thermodynamic,tao2018direct, yonezawa2019nematic},  Sr$_x$Bi$_2$Se$_3$~\cite{wang2019evidence, kostylev2020uniaxial}, and Nb$_x$Bi$_2$Se$_3$~\cite{shen2017nematic,asaba2017rotational},
superconducting topological semimetal PbTaSe$_2$~\cite{XinLu2020evidence}, and more recently  twisted bilayer graphene~\cite{kerelsky2019maximized, cao2020nematicity}, we believe that our results pave an important step toward experimentally realizing charge-$4e$ phases in quantum materials. 
Our analysis can be applied to the nematic superconducting ordering of twisted bilayer graphene~\cite{chichinadze2020nematic, lothman2021nematic}, though the pairing mechanism in the twisted bilayer graphene has not been completely understood
~\cite{sherkunov2018electronic, gonzalez2019kohn, lian2019twisted, fischer2021spin}.
A possible experiment to detect the charge-$4e$ superconductivity is to measure the magnetic flux quantization, which is half of the quantization value in an ordinary charge-$2e$ superconductor.

{\it Acknowledgement.---} We thank Wen Huang for helpful discussions. This work is supported in part by the NSFC under Grant No. 11825404 (HY), the MOSTC under Grant No. 2018YFA0305604 (H.Y.), Beijing Natural Science Foundation under Grant No. Z180010 (H.Y.), the Strategic Priority Research Program of Chinese Academy of Sciences under Grant No. XDB28000000 (H.Y.), and Beijing Municipal Science and Technology Commission under Grant No. Z181100004218001 (H.Y.). SKJ is supported by the Simons Foundation via the It From Qubit Collaboration. 

{\it Note added:} While the present paper was close to be completed, we notice an interesting work on a similar topic in Ref.~\cite{fernandes2021charge}. 
Our results of Ginzburg-Landau theory analysis in three dimensions are qualitatively consistent with ones in Ref.~\cite{fernandes2021charge}.  
We additionally analyzed the topological defect theory in two dimensions which supports the possibility of realizing charge-4e phase. 

\bibliography{charge4e.bib}

\begin{thebibliography}{52}%
\makeatletter
\providecommand \@ifxundefined [1]{%
 \@ifx{#1\undefined}
}%
\providecommand \@ifnum [1]{%
 \ifnum #1\expandafter \@firstoftwo
 \else \expandafter \@secondoftwo
 \fi
}%
\providecommand \@ifx [1]{%
 \ifx #1\expandafter \@firstoftwo
 \else \expandafter \@secondoftwo
 \fi
}%
\providecommand \natexlab [1]{#1}%
\providecommand \enquote  [1]{``#1''}%
\providecommand \bibnamefont  [1]{#1}%
\providecommand \bibfnamefont [1]{#1}%
\providecommand \citenamefont [1]{#1}%
\providecommand \href@noop [0]{\@secondoftwo}%
\providecommand \href [0]{\begingroup \@sanitize@url \@href}%
\providecommand \@href[1]{\@@startlink{#1}\@@href}%
\providecommand \@@href[1]{\endgroup#1\@@endlink}%
\providecommand \@sanitize@url [0]{\catcode `\\12\catcode `\$12\catcode
  `\&12\catcode `\#12\catcode `\^12\catcode `\_12\catcode `\%12\relax}%
\providecommand \@@startlink[1]{}%
\providecommand \@@endlink[0]{}%
\providecommand \url  [0]{\begingroup\@sanitize@url \@url }%
\providecommand \@url [1]{\endgroup\@href {#1}{\urlprefix }}%
\providecommand \urlprefix  [0]{URL }%
\providecommand \Eprint [0]{\href }%
\providecommand \doibase [0]{https://doi.org/}%
\providecommand \selectlanguage [0]{\@gobble}%
\providecommand \bibinfo  [0]{\@secondoftwo}%
\providecommand \bibfield  [0]{\@secondoftwo}%
\providecommand \translation [1]{[#1]}%
\providecommand \BibitemOpen [0]{}%
\providecommand \bibitemStop [0]{}%
\providecommand \bibitemNoStop [0]{.\EOS\space}%
\providecommand \EOS [0]{\spacefactor3000\relax}%
\providecommand \BibitemShut  [1]{\csname bibitem#1\endcsname}%
\let\auto@bib@innerbib\@empty
\bibitem [{\citenamefont {Korshunov}(1985)}]{korshunov1985two}%
  \BibitemOpen
  \bibfield  {author} {\bibinfo {author} {\bibfnamefont {S.}~\bibnamefont
  {Korshunov}},\ }\href@noop {} {\bibfield  {journal} {\bibinfo  {journal} {Zh.
  Eksp. Teor. Fiz}\ }\textbf {\bibinfo {volume} {89}},\ \bibinfo {pages} {539}
  (\bibinfo {year} {1985})}\BibitemShut {NoStop}%
\bibitem [{\citenamefont {Kivelson}\ \emph {et~al.}(1990)\citenamefont
  {Kivelson}, \citenamefont {Emery},\ and\ \citenamefont
  {Lin}}]{kivelson1990doped}%
  \BibitemOpen
  \bibfield  {author} {\bibinfo {author} {\bibfnamefont {S.}~\bibnamefont
  {Kivelson}}, \bibinfo {author} {\bibfnamefont {V.}~\bibnamefont {Emery}},\
  and\ \bibinfo {author} {\bibfnamefont {H.}~\bibnamefont {Lin}},\ }\href@noop
  {} {\bibfield  {journal} {\bibinfo  {journal} {Phys. Rev. B}\ }\textbf
  {\bibinfo {volume} {42}},\ \bibinfo {pages} {6523} (\bibinfo {year}
  {1990})}\BibitemShut {NoStop}%
\bibitem [{\citenamefont {Wu}(2005)}]{wu2005competing}%
  \BibitemOpen
  \bibfield  {author} {\bibinfo {author} {\bibfnamefont {C.}~\bibnamefont
  {Wu}},\ }\href@noop {} {\bibfield  {journal} {\bibinfo  {journal} {Phys. Rev.
  Lett.}\ }\textbf {\bibinfo {volume} {95}},\ \bibinfo {pages} {266404}
  (\bibinfo {year} {2005})}\BibitemShut {NoStop}%
\bibitem [{\citenamefont {Berg}\ \emph
  {et~al.}(2009{\natexlab{a}})\citenamefont {Berg}, \citenamefont {Fradkin},\
  and\ \citenamefont {Kivelson}}]{berg2009charge}%
  \BibitemOpen
  \bibfield  {author} {\bibinfo {author} {\bibfnamefont {E.}~\bibnamefont
  {Berg}}, \bibinfo {author} {\bibfnamefont {E.}~\bibnamefont {Fradkin}},\ and\
  \bibinfo {author} {\bibfnamefont {S.~A.}\ \bibnamefont {Kivelson}},\
  }\href@noop {} {\bibfield  {journal} {\bibinfo  {journal} {Nature Physics}\
  }\textbf {\bibinfo {volume} {5}},\ \bibinfo {pages} {830} (\bibinfo {year}
  {2009}{\natexlab{a}})}\BibitemShut {NoStop}%
\bibitem [{\citenamefont {Jiang}\ \emph {et~al.}(2017)\citenamefont {Jiang},
  \citenamefont {Li}, \citenamefont {Kivelson},\ and\ \citenamefont
  {Yao}}]{jiang2017charge}%
  \BibitemOpen
  \bibfield  {author} {\bibinfo {author} {\bibfnamefont {Y.-F.}\ \bibnamefont
  {Jiang}}, \bibinfo {author} {\bibfnamefont {Z.-X.}\ \bibnamefont {Li}},
  \bibinfo {author} {\bibfnamefont {S.~A.}\ \bibnamefont {Kivelson}},\ and\
  \bibinfo {author} {\bibfnamefont {H.}~\bibnamefont {Yao}},\ }\href@noop {}
  {\bibfield  {journal} {\bibinfo  {journal} {Phys. Rev. B}\ }\textbf {\bibinfo
  {volume} {95}},\ \bibinfo {pages} {241103} (\bibinfo {year}
  {2017})}\BibitemShut {NoStop}%
\bibitem [{Note1()}]{Note1}%
  \BibitemOpen
  \bibinfo {note} {In this paper, we refer the charge-$2e$ superconductivity as
  conventional superconductivity to distinguish it from the charge-$4e$
  superconductivity. For example, both $s$-wave isotropic superconductor and
  $d$-wave nematic superconductor are conventional in this
  context.}\BibitemShut {Stop}%
\bibitem [{\citenamefont {Fulde}\ and\ \citenamefont
  {Ferrell}(1964)}]{fulde1964superconductivity}%
  \BibitemOpen
  \bibfield  {author} {\bibinfo {author} {\bibfnamefont {P.}~\bibnamefont
  {Fulde}}\ and\ \bibinfo {author} {\bibfnamefont {R.~A.}\ \bibnamefont
  {Ferrell}},\ }\href@noop {} {\bibfield  {journal} {\bibinfo  {journal} {Phys.
  Rev.}\ }\textbf {\bibinfo {volume} {135}},\ \bibinfo {pages} {A550} (\bibinfo
  {year} {1964})}\BibitemShut {NoStop}%
\bibitem [{\citenamefont {Larkin}\ and\ \citenamefont
  {Ovchinnikov}(1965)}]{larkin1965zh}%
  \BibitemOpen
  \bibfield  {author} {\bibinfo {author} {\bibfnamefont {A.}~\bibnamefont
  {Larkin}}\ and\ \bibinfo {author} {\bibfnamefont {Y.~N.}\ \bibnamefont
  {Ovchinnikov}},\ }\href@noop {} {\bibfield  {journal} {\bibinfo  {journal}
  {JETP}\ }\textbf {\bibinfo {volume} {20}},\ \bibinfo {pages} {762} (\bibinfo
  {year} {1965})}\BibitemShut {NoStop}%
\bibitem [{\citenamefont {Radzihovsky}\ and\ \citenamefont
  {Vishwanath}(2009)}]{radzihovsky2009quantum}%
  \BibitemOpen
  \bibfield  {author} {\bibinfo {author} {\bibfnamefont {L.}~\bibnamefont
  {Radzihovsky}}\ and\ \bibinfo {author} {\bibfnamefont {A.}~\bibnamefont
  {Vishwanath}},\ }\href@noop {} {\bibfield  {journal} {\bibinfo  {journal}
  {Phys. Rev. Lett.}\ }\textbf {\bibinfo {volume} {103}},\ \bibinfo {pages}
  {010404} (\bibinfo {year} {2009})}\BibitemShut {NoStop}%
\bibitem [{\citenamefont {Agterberg}\ and\ \citenamefont
  {Tsunetsugu}(2008)}]{Agterberg2008NPdislocations}%
  \BibitemOpen
  \bibfield  {author} {\bibinfo {author} {\bibfnamefont {D.}~\bibnamefont
  {Agterberg}}\ and\ \bibinfo {author} {\bibfnamefont {H.}~\bibnamefont
  {Tsunetsugu}},\ }\href@noop {} {\bibfield  {journal} {\bibinfo  {journal}
  {Nature Physics}\ }\textbf {\bibinfo {volume} {4}},\ \bibinfo {pages} {639}
  (\bibinfo {year} {2008})}\BibitemShut {NoStop}%
\bibitem [{\citenamefont {Wang}\ \emph {et~al.}(2010)\citenamefont {Wang},
  \citenamefont {Gao}, \citenamefont {Jian},\ and\ \citenamefont
  {Zhai}}]{wang2010spin}%
  \BibitemOpen
  \bibfield  {author} {\bibinfo {author} {\bibfnamefont {C.}~\bibnamefont
  {Wang}}, \bibinfo {author} {\bibfnamefont {C.}~\bibnamefont {Gao}}, \bibinfo
  {author} {\bibfnamefont {C.-M.}\ \bibnamefont {Jian}},\ and\ \bibinfo
  {author} {\bibfnamefont {H.}~\bibnamefont {Zhai}},\ }\href@noop {} {\bibfield
   {journal} {\bibinfo  {journal} {Phys. Rev. Lett.}\ }\textbf {\bibinfo
  {volume} {105}},\ \bibinfo {pages} {160403} (\bibinfo {year}
  {2010})}\BibitemShut {NoStop}%
\bibitem [{\citenamefont {Cho}\ \emph {et~al.}(2012)\citenamefont {Cho},
  \citenamefont {Bardarson}, \citenamefont {Lu},\ and\ \citenamefont
  {Moore}}]{cho2012superconductivity}%
  \BibitemOpen
  \bibfield  {author} {\bibinfo {author} {\bibfnamefont {G.~Y.}\ \bibnamefont
  {Cho}}, \bibinfo {author} {\bibfnamefont {J.~H.}\ \bibnamefont {Bardarson}},
  \bibinfo {author} {\bibfnamefont {Y.-M.}\ \bibnamefont {Lu}},\ and\ \bibinfo
  {author} {\bibfnamefont {J.~E.}\ \bibnamefont {Moore}},\ }\href@noop {}
  {\bibfield  {journal} {\bibinfo  {journal} {Phys. Rev. B}\ }\textbf {\bibinfo
  {volume} {86}},\ \bibinfo {pages} {214514} (\bibinfo {year}
  {2012})}\BibitemShut {NoStop}%
\bibitem [{\citenamefont {Lee}(2014)}]{lee2014amperean}%
  \BibitemOpen
  \bibfield  {author} {\bibinfo {author} {\bibfnamefont {P.~A.}\ \bibnamefont
  {Lee}},\ }\href@noop {} {\bibfield  {journal} {\bibinfo  {journal} {Phys.
  Rev. X}\ }\textbf {\bibinfo {volume} {4}},\ \bibinfo {pages} {031017}
  (\bibinfo {year} {2014})}\BibitemShut {NoStop}%
\bibitem [{\citenamefont {Maciejko}\ and\ \citenamefont
  {Nandkishore}(2014)}]{maciejko2014weyl}%
  \BibitemOpen
  \bibfield  {author} {\bibinfo {author} {\bibfnamefont {J.}~\bibnamefont
  {Maciejko}}\ and\ \bibinfo {author} {\bibfnamefont {R.}~\bibnamefont
  {Nandkishore}},\ }\href@noop {} {\bibfield  {journal} {\bibinfo  {journal}
  {Phys. Rev. B}\ }\textbf {\bibinfo {volume} {90}},\ \bibinfo {pages} {035126}
  (\bibinfo {year} {2014})}\BibitemShut {NoStop}%
\bibitem [{\citenamefont {Jian}\ \emph {et~al.}(2015)\citenamefont {Jian},
  \citenamefont {Jiang},\ and\ \citenamefont {Yao}}]{jian2015emergent}%
  \BibitemOpen
  \bibfield  {author} {\bibinfo {author} {\bibfnamefont {S.-K.}\ \bibnamefont
  {Jian}}, \bibinfo {author} {\bibfnamefont {Y.-F.}\ \bibnamefont {Jiang}},\
  and\ \bibinfo {author} {\bibfnamefont {H.}~\bibnamefont {Yao}},\ }\href@noop
  {} {\bibfield  {journal} {\bibinfo  {journal} {Phys. Rev. Lett.}\ }\textbf
  {\bibinfo {volume} {114}},\ \bibinfo {pages} {237001} (\bibinfo {year}
  {2015})}\BibitemShut {NoStop}%
\bibitem [{\citenamefont {Jian}\ \emph {et~al.}(2017)\citenamefont {Jian},
  \citenamefont {Lin}, \citenamefont {Maciejko},\ and\ \citenamefont
  {Yao}}]{jian2017emergence}%
  \BibitemOpen
  \bibfield  {author} {\bibinfo {author} {\bibfnamefont {S.-K.}\ \bibnamefont
  {Jian}}, \bibinfo {author} {\bibfnamefont {C.-H.}\ \bibnamefont {Lin}},
  \bibinfo {author} {\bibfnamefont {J.}~\bibnamefont {Maciejko}},\ and\
  \bibinfo {author} {\bibfnamefont {H.}~\bibnamefont {Yao}},\ }\href@noop {}
  {\bibfield  {journal} {\bibinfo  {journal} {Phys. Rev. Lett.}\ }\textbf
  {\bibinfo {volume} {118}},\ \bibinfo {pages} {166802} (\bibinfo {year}
  {2017})}\BibitemShut {NoStop}%
\bibitem [{\citenamefont {Jian}\ \emph {et~al.}(2020)\citenamefont {Jian},
  \citenamefont {Scherer},\ and\ \citenamefont {Yao}}]{jian2020mass}%
  \BibitemOpen
  \bibfield  {author} {\bibinfo {author} {\bibfnamefont {S.-K.}\ \bibnamefont
  {Jian}}, \bibinfo {author} {\bibfnamefont {M.~M.}\ \bibnamefont {Scherer}},\
  and\ \bibinfo {author} {\bibfnamefont {H.}~\bibnamefont {Yao}},\ }\href@noop
  {} {\bibfield  {journal} {\bibinfo  {journal} {Phys. Rev. Research}\ }\textbf
  {\bibinfo {volume} {2}},\ \bibinfo {pages} {013034} (\bibinfo {year}
  {2020})}\BibitemShut {NoStop}%
\bibitem [{\citenamefont {Han}\ \emph {et~al.}(2020)\citenamefont {Han},
  \citenamefont {Kivelson},\ and\ \citenamefont {Yao}}]{Han2020PRL-PDW}%
  \BibitemOpen
  \bibfield  {author} {\bibinfo {author} {\bibfnamefont {Z.}~\bibnamefont
  {Han}}, \bibinfo {author} {\bibfnamefont {S.~A.}\ \bibnamefont {Kivelson}},\
  and\ \bibinfo {author} {\bibfnamefont {H.}~\bibnamefont {Yao}},\ }\href@noop
  {} {\bibfield  {journal} {\bibinfo  {journal} {Phys. Rev. Lett.}\ }\textbf
  {\bibinfo {volume} {125}},\ \bibinfo {pages} {167001} (\bibinfo {year}
  {2020})}\BibitemShut {NoStop}%
\bibitem [{\citenamefont {Agterberg}\ \emph {et~al.}(2020)\citenamefont
  {Agterberg}, \citenamefont {Davis}, \citenamefont {Edkins}, \citenamefont
  {Fradkin}, \citenamefont {Van~Harlingen}, \citenamefont {Kivelson},
  \citenamefont {Lee}, \citenamefont {Radzihovsky}, \citenamefont {Tranquada},\
  and\ \citenamefont {Wang}}]{kivelson2020-arcmp}%
  \BibitemOpen
  \bibfield  {author} {\bibinfo {author} {\bibfnamefont {D.~F.}\ \bibnamefont
  {Agterberg}}, \bibinfo {author} {\bibfnamefont {J.~S.}\ \bibnamefont
  {Davis}}, \bibinfo {author} {\bibfnamefont {S.~D.}\ \bibnamefont {Edkins}},
  \bibinfo {author} {\bibfnamefont {E.}~\bibnamefont {Fradkin}}, \bibinfo
  {author} {\bibfnamefont {D.~J.}\ \bibnamefont {Van~Harlingen}}, \bibinfo
  {author} {\bibfnamefont {S.~A.}\ \bibnamefont {Kivelson}}, \bibinfo {author}
  {\bibfnamefont {P.~A.}\ \bibnamefont {Lee}}, \bibinfo {author} {\bibfnamefont
  {L.}~\bibnamefont {Radzihovsky}}, \bibinfo {author} {\bibfnamefont {J.~M.}\
  \bibnamefont {Tranquada}},\ and\ \bibinfo {author} {\bibfnamefont
  {Y.}~\bibnamefont {Wang}},\ }\href@noop {} {\bibfield  {journal} {\bibinfo
  {journal} {Annual Review of Condensed Matter Physics}\ }\textbf {\bibinfo
  {volume} {11}},\ \bibinfo {pages} {231} (\bibinfo {year} {2020})}\BibitemShut
  {NoStop}%
\bibitem [{\citenamefont {Chung}\ \emph {et~al.}(2012)\citenamefont {Chung},
  \citenamefont {Raghu}, \citenamefont {Kapitulnik},\ and\ \citenamefont
  {Kivelson}}]{chung2012charge}%
  \BibitemOpen
  \bibfield  {author} {\bibinfo {author} {\bibfnamefont {S.~B.}\ \bibnamefont
  {Chung}}, \bibinfo {author} {\bibfnamefont {S.}~\bibnamefont {Raghu}},
  \bibinfo {author} {\bibfnamefont {A.}~\bibnamefont {Kapitulnik}},\ and\
  \bibinfo {author} {\bibfnamefont {S.~A.}\ \bibnamefont {Kivelson}},\
  }\href@noop {} {\bibfield  {journal} {\bibinfo  {journal} {Physical Review
  B}\ }\textbf {\bibinfo {volume} {86}},\ \bibinfo {pages} {064525} (\bibinfo
  {year} {2012})}\BibitemShut {NoStop}%
\bibitem [{\citenamefont {Moon}(2012)}]{Moon2012PRB-charge4e}%
  \BibitemOpen
  \bibfield  {author} {\bibinfo {author} {\bibfnamefont {E.-G.}\ \bibnamefont
  {Moon}},\ }\href@noop {} {\bibfield  {journal} {\bibinfo  {journal} {Phys.
  Rev. B}\ }\textbf {\bibinfo {volume} {85}},\ \bibinfo {pages} {245123}
  (\bibinfo {year} {2012})}\BibitemShut {NoStop}%
\bibitem [{\citenamefont {Xu}\ and\ \citenamefont
  {Balents}(2018)}]{xu2018topological}%
  \BibitemOpen
  \bibfield  {author} {\bibinfo {author} {\bibfnamefont {C.}~\bibnamefont
  {Xu}}\ and\ \bibinfo {author} {\bibfnamefont {L.}~\bibnamefont {Balents}},\
  }\href@noop {} {\bibfield  {journal} {\bibinfo  {journal} {Physical review
  letters}\ }\textbf {\bibinfo {volume} {121}},\ \bibinfo {pages} {087001}
  (\bibinfo {year} {2018})}\BibitemShut {NoStop}%
\bibitem [{\citenamefont {Takeuchi}(2021)}]{takeuchi2020phase}%
  \BibitemOpen
  \bibfield  {author} {\bibinfo {author} {\bibfnamefont {H.}~\bibnamefont
  {Takeuchi}},\ }\href@noop {} {\bibfield  {journal} {\bibinfo  {journal}
  {Phys. Rev. A}\ }\textbf {\bibinfo {volume} {104}},\ \bibinfo {pages}
  {013316} (\bibinfo {year} {2021})}\BibitemShut {NoStop}%
\bibitem [{\citenamefont {Fujita}\ \emph {et~al.}(2004)\citenamefont {Fujita},
  \citenamefont {Goka}, \citenamefont {Yamada}, \citenamefont {Tranquada},\
  and\ \citenamefont {Regnault}}]{fujita2004stripe}%
  \BibitemOpen
  \bibfield  {author} {\bibinfo {author} {\bibfnamefont {M.}~\bibnamefont
  {Fujita}}, \bibinfo {author} {\bibfnamefont {H.}~\bibnamefont {Goka}},
  \bibinfo {author} {\bibfnamefont {K.}~\bibnamefont {Yamada}}, \bibinfo
  {author} {\bibfnamefont {J.}~\bibnamefont {Tranquada}},\ and\ \bibinfo
  {author} {\bibfnamefont {L.}~\bibnamefont {Regnault}},\ }\href@noop {}
  {\bibfield  {journal} {\bibinfo  {journal} {Phys. Rev. B}\ }\textbf {\bibinfo
  {volume} {70}},\ \bibinfo {pages} {104517} (\bibinfo {year}
  {2004})}\BibitemShut {NoStop}%
\bibitem [{\citenamefont {Tranquada}\ \emph {et~al.}(2008)\citenamefont
  {Tranquada}, \citenamefont {Gu}, \citenamefont {H\"ucker}, \citenamefont
  {Jie}, \citenamefont {Kang}, \citenamefont {Klingeler}, \citenamefont {Li},
  \citenamefont {Tristan}, \citenamefont {Wen}, \citenamefont {Xu},
  \citenamefont {Xu}, \citenamefont {Zhou},\ and\ \citenamefont
  {v.~Zimmermann}}]{tranquada2008evidence}%
  \BibitemOpen
  \bibfield  {author} {\bibinfo {author} {\bibfnamefont {J.~M.}\ \bibnamefont
  {Tranquada}}, \bibinfo {author} {\bibfnamefont {G.~D.}\ \bibnamefont {Gu}},
  \bibinfo {author} {\bibfnamefont {M.}~\bibnamefont {H\"ucker}}, \bibinfo
  {author} {\bibfnamefont {Q.}~\bibnamefont {Jie}}, \bibinfo {author}
  {\bibfnamefont {H.-J.}\ \bibnamefont {Kang}}, \bibinfo {author}
  {\bibfnamefont {R.}~\bibnamefont {Klingeler}}, \bibinfo {author}
  {\bibfnamefont {Q.}~\bibnamefont {Li}}, \bibinfo {author} {\bibfnamefont
  {N.}~\bibnamefont {Tristan}}, \bibinfo {author} {\bibfnamefont {J.~S.}\
  \bibnamefont {Wen}}, \bibinfo {author} {\bibfnamefont {G.~Y.}\ \bibnamefont
  {Xu}}, \bibinfo {author} {\bibfnamefont {Z.~J.}\ \bibnamefont {Xu}}, \bibinfo
  {author} {\bibfnamefont {J.}~\bibnamefont {Zhou}},\ and\ \bibinfo {author}
  {\bibfnamefont {M.}~\bibnamefont {v.~Zimmermann}},\ }\href@noop {} {\bibfield
   {journal} {\bibinfo  {journal} {Phys. Rev. B}\ }\textbf {\bibinfo {volume}
  {78}},\ \bibinfo {pages} {174529} (\bibinfo {year} {2008})}\BibitemShut
  {NoStop}%
\bibitem [{\citenamefont {H{\"u}cker}\ \emph {et~al.}(2011)\citenamefont
  {H{\"u}cker}, \citenamefont {Zimmermann}, \citenamefont {Gu}, \citenamefont
  {Xu}, \citenamefont {Wen}, \citenamefont {Xu}, \citenamefont {Kang},
  \citenamefont {Zheludev},\ and\ \citenamefont
  {Tranquada}}]{hucker2011stripe}%
  \BibitemOpen
  \bibfield  {author} {\bibinfo {author} {\bibfnamefont {M.}~\bibnamefont
  {H{\"u}cker}}, \bibinfo {author} {\bibfnamefont {M.~v.}\ \bibnamefont
  {Zimmermann}}, \bibinfo {author} {\bibfnamefont {G.}~\bibnamefont {Gu}},
  \bibinfo {author} {\bibfnamefont {Z.}~\bibnamefont {Xu}}, \bibinfo {author}
  {\bibfnamefont {J.}~\bibnamefont {Wen}}, \bibinfo {author} {\bibfnamefont
  {G.}~\bibnamefont {Xu}}, \bibinfo {author} {\bibfnamefont {H.}~\bibnamefont
  {Kang}}, \bibinfo {author} {\bibfnamefont {A.}~\bibnamefont {Zheludev}},\
  and\ \bibinfo {author} {\bibfnamefont {J.~M.}\ \bibnamefont {Tranquada}},\
  }\href@noop {} {\bibfield  {journal} {\bibinfo  {journal} {Phys. Rev. B}\
  }\textbf {\bibinfo {volume} {83}},\ \bibinfo {pages} {104506} (\bibinfo
  {year} {2011})}\BibitemShut {NoStop}%
\bibitem [{\citenamefont {Berg}\ \emph {et~al.}(2007)\citenamefont {Berg},
  \citenamefont {Fradkin}, \citenamefont {Kim}, \citenamefont {Kivelson},
  \citenamefont {Oganesyan}, \citenamefont {Tranquada},\ and\ \citenamefont
  {Zhang}}]{berg2007dynamical}%
  \BibitemOpen
  \bibfield  {author} {\bibinfo {author} {\bibfnamefont {E.}~\bibnamefont
  {Berg}}, \bibinfo {author} {\bibfnamefont {E.}~\bibnamefont {Fradkin}},
  \bibinfo {author} {\bibfnamefont {E.-A.}\ \bibnamefont {Kim}}, \bibinfo
  {author} {\bibfnamefont {S.~A.}\ \bibnamefont {Kivelson}}, \bibinfo {author}
  {\bibfnamefont {V.}~\bibnamefont {Oganesyan}}, \bibinfo {author}
  {\bibfnamefont {J.~M.}\ \bibnamefont {Tranquada}},\ and\ \bibinfo {author}
  {\bibfnamefont {S.-C.}\ \bibnamefont {Zhang}},\ }\href@noop {} {\bibfield
  {journal} {\bibinfo  {journal} {Phys. Rev. Lett.}\ }\textbf {\bibinfo
  {volume} {99}},\ \bibinfo {pages} {127003} (\bibinfo {year}
  {2007})}\BibitemShut {NoStop}%
\bibitem [{\citenamefont {Berg}\ \emph
  {et~al.}(2009{\natexlab{b}})\citenamefont {Berg}, \citenamefont {Fradkin},\
  and\ \citenamefont {Kivelson}}]{berg2009theory}%
  \BibitemOpen
  \bibfield  {author} {\bibinfo {author} {\bibfnamefont {E.}~\bibnamefont
  {Berg}}, \bibinfo {author} {\bibfnamefont {E.}~\bibnamefont {Fradkin}},\ and\
  \bibinfo {author} {\bibfnamefont {S.~A.}\ \bibnamefont {Kivelson}},\
  }\href@noop {} {\bibfield  {journal} {\bibinfo  {journal} {Phys. Rev. B}\
  }\textbf {\bibinfo {volume} {79}},\ \bibinfo {pages} {064515} (\bibinfo
  {year} {2009}{\natexlab{b}})}\BibitemShut {NoStop}%
\bibitem [{\citenamefont {Matano}\ \emph {et~al.}(2016)\citenamefont {Matano},
  \citenamefont {Kriener}, \citenamefont {Segawa}, \citenamefont {Ando},\ and\
  \citenamefont {Zheng}}]{matano2016spin}%
  \BibitemOpen
  \bibfield  {author} {\bibinfo {author} {\bibfnamefont {K.}~\bibnamefont
  {Matano}}, \bibinfo {author} {\bibfnamefont {M.}~\bibnamefont {Kriener}},
  \bibinfo {author} {\bibfnamefont {K.}~\bibnamefont {Segawa}}, \bibinfo
  {author} {\bibfnamefont {Y.}~\bibnamefont {Ando}},\ and\ \bibinfo {author}
  {\bibfnamefont {G.-q.}\ \bibnamefont {Zheng}},\ }\href@noop {} {\bibfield
  {journal} {\bibinfo  {journal} {Nature Physics}\ }\textbf {\bibinfo {volume}
  {12}},\ \bibinfo {pages} {852} (\bibinfo {year} {2016})}\BibitemShut
  {NoStop}%
\bibitem [{\citenamefont {Yonezawa}\ \emph {et~al.}(2017)\citenamefont
  {Yonezawa}, \citenamefont {Tajiri}, \citenamefont {Nakata}, \citenamefont
  {Nagai}, \citenamefont {Wang}, \citenamefont {Segawa}, \citenamefont {Ando},\
  and\ \citenamefont {Maeno}}]{yonezawa2017thermodynamic}%
  \BibitemOpen
  \bibfield  {author} {\bibinfo {author} {\bibfnamefont {S.}~\bibnamefont
  {Yonezawa}}, \bibinfo {author} {\bibfnamefont {K.}~\bibnamefont {Tajiri}},
  \bibinfo {author} {\bibfnamefont {S.}~\bibnamefont {Nakata}}, \bibinfo
  {author} {\bibfnamefont {Y.}~\bibnamefont {Nagai}}, \bibinfo {author}
  {\bibfnamefont {Z.}~\bibnamefont {Wang}}, \bibinfo {author} {\bibfnamefont
  {K.}~\bibnamefont {Segawa}}, \bibinfo {author} {\bibfnamefont
  {Y.}~\bibnamefont {Ando}},\ and\ \bibinfo {author} {\bibfnamefont
  {Y.}~\bibnamefont {Maeno}},\ }\href@noop {} {\bibfield  {journal} {\bibinfo
  {journal} {Nature Physics}\ }\textbf {\bibinfo {volume} {13}},\ \bibinfo
  {pages} {123} (\bibinfo {year} {2017})}\BibitemShut {NoStop}%
\bibitem [{\citenamefont {Tao}\ \emph {et~al.}(2018)\citenamefont {Tao},
  \citenamefont {Yan}, \citenamefont {Liu}, \citenamefont {Wang}, \citenamefont
  {Ando}, \citenamefont {Wang}, \citenamefont {Zhang},\ and\ \citenamefont
  {Feng}}]{tao2018direct}%
  \BibitemOpen
  \bibfield  {author} {\bibinfo {author} {\bibfnamefont {R.}~\bibnamefont
  {Tao}}, \bibinfo {author} {\bibfnamefont {Y.-J.}\ \bibnamefont {Yan}},
  \bibinfo {author} {\bibfnamefont {X.}~\bibnamefont {Liu}}, \bibinfo {author}
  {\bibfnamefont {Z.-W.}\ \bibnamefont {Wang}}, \bibinfo {author}
  {\bibfnamefont {Y.}~\bibnamefont {Ando}}, \bibinfo {author} {\bibfnamefont
  {Q.-H.}\ \bibnamefont {Wang}}, \bibinfo {author} {\bibfnamefont
  {T.}~\bibnamefont {Zhang}},\ and\ \bibinfo {author} {\bibfnamefont {D.-L.}\
  \bibnamefont {Feng}},\ }\href@noop {} {\bibfield  {journal} {\bibinfo
  {journal} {Phys. Rev. X}\ }\textbf {\bibinfo {volume} {8}},\ \bibinfo {pages}
  {041024} (\bibinfo {year} {2018})}\BibitemShut {NoStop}%
\bibitem [{\citenamefont {Yonezawa}(2019)}]{yonezawa2019nematic}%
  \BibitemOpen
  \bibfield  {author} {\bibinfo {author} {\bibfnamefont {S.}~\bibnamefont
  {Yonezawa}},\ }\href@noop {} {\bibfield  {journal} {\bibinfo  {journal}
  {Condensed Matter}\ }\textbf {\bibinfo {volume} {4}},\ \bibinfo {pages} {2}
  (\bibinfo {year} {2019})}\BibitemShut {NoStop}%
\bibitem [{\citenamefont {Fu}(2014)}]{fu2014odd}%
  \BibitemOpen
  \bibfield  {author} {\bibinfo {author} {\bibfnamefont {L.}~\bibnamefont
  {Fu}},\ }\href@noop {} {\bibfield  {journal} {\bibinfo  {journal} {Phys. Rev.
  B}\ }\textbf {\bibinfo {volume} {90}},\ \bibinfo {pages} {100509} (\bibinfo
  {year} {2014})}\BibitemShut {NoStop}%
\bibitem [{\citenamefont {Venderbos}\ \emph {et~al.}(2016)\citenamefont
  {Venderbos}, \citenamefont {Kozii},\ and\ \citenamefont
  {Fu}}]{venderbos2016odd}%
  \BibitemOpen
  \bibfield  {author} {\bibinfo {author} {\bibfnamefont {J.~W.}\ \bibnamefont
  {Venderbos}}, \bibinfo {author} {\bibfnamefont {V.}~\bibnamefont {Kozii}},\
  and\ \bibinfo {author} {\bibfnamefont {L.}~\bibnamefont {Fu}},\ }\href@noop
  {} {\bibfield  {journal} {\bibinfo  {journal} {Phys. Rev. B}\ }\textbf
  {\bibinfo {volume} {94}},\ \bibinfo {pages} {180504} (\bibinfo {year}
  {2016})}\BibitemShut {NoStop}%
\bibitem [{\citenamefont {Hecker}\ and\ \citenamefont
  {Schmalian}(2018)}]{hecker2018vestigial}%
  \BibitemOpen
  \bibfield  {author} {\bibinfo {author} {\bibfnamefont {M.}~\bibnamefont
  {Hecker}}\ and\ \bibinfo {author} {\bibfnamefont {J.}~\bibnamefont
  {Schmalian}},\ }\href@noop {} {\bibfield  {journal} {\bibinfo  {journal} {npj
  Quantum Materials}\ }\textbf {\bibinfo {volume} {3}},\ \bibinfo {pages} {26}
  (\bibinfo {year} {2018})}\BibitemShut {NoStop}%
\bibitem [{\citenamefont {Wang}\ \emph {et~al.}(2019)\citenamefont {Wang},
  \citenamefont {Ran}, \citenamefont {Li}, \citenamefont {Ma}, \citenamefont
  {Bao}, \citenamefont {Cai}, \citenamefont {Zhang}, \citenamefont {Nakajima},
  \citenamefont {Ohira-Kawamura}, \citenamefont {{\v{C}}erm{\'a}k},
  \citenamefont {Schneidewind}, \citenamefont {Savrasov}, \citenamefont {Wan},\
  and\ \citenamefont {Wen}}]{wang2019evidence}%
  \BibitemOpen
  \bibfield  {author} {\bibinfo {author} {\bibfnamefont {J.}~\bibnamefont
  {Wang}}, \bibinfo {author} {\bibfnamefont {K.}~\bibnamefont {Ran}}, \bibinfo
  {author} {\bibfnamefont {S.}~\bibnamefont {Li}}, \bibinfo {author}
  {\bibfnamefont {Z.}~\bibnamefont {Ma}}, \bibinfo {author} {\bibfnamefont
  {S.}~\bibnamefont {Bao}}, \bibinfo {author} {\bibfnamefont {Z.}~\bibnamefont
  {Cai}}, \bibinfo {author} {\bibfnamefont {Y.}~\bibnamefont {Zhang}}, \bibinfo
  {author} {\bibfnamefont {K.}~\bibnamefont {Nakajima}}, \bibinfo {author}
  {\bibfnamefont {S.}~\bibnamefont {Ohira-Kawamura}}, \bibinfo {author}
  {\bibfnamefont {P.}~\bibnamefont {{\v{C}}erm{\'a}k}}, \bibinfo {author}
  {\bibfnamefont {A.}~\bibnamefont {Schneidewind}}, \bibinfo {author}
  {\bibfnamefont {S.~Y.}\ \bibnamefont {Savrasov}}, \bibinfo {author}
  {\bibfnamefont {X.}~\bibnamefont {Wan}},\ and\ \bibinfo {author}
  {\bibfnamefont {J.}~\bibnamefont {Wen}},\ }\href@noop {} {\bibfield
  {journal} {\bibinfo  {journal} {Nature communications}\ }\textbf {\bibinfo
  {volume} {10}},\ \bibinfo {pages} {2802} (\bibinfo {year}
  {2019})}\BibitemShut {NoStop}%
\bibitem [{\citenamefont {Kostylev}\ \emph {et~al.}(2020)\citenamefont
  {Kostylev}, \citenamefont {Yonezawa}, \citenamefont {Wang}, \citenamefont
  {Ando},\ and\ \citenamefont {Maeno}}]{kostylev2020uniaxial}%
  \BibitemOpen
  \bibfield  {author} {\bibinfo {author} {\bibfnamefont {I.}~\bibnamefont
  {Kostylev}}, \bibinfo {author} {\bibfnamefont {S.}~\bibnamefont {Yonezawa}},
  \bibinfo {author} {\bibfnamefont {Z.}~\bibnamefont {Wang}}, \bibinfo {author}
  {\bibfnamefont {Y.}~\bibnamefont {Ando}},\ and\ \bibinfo {author}
  {\bibfnamefont {Y.}~\bibnamefont {Maeno}},\ }\href@noop {} {\bibfield
  {journal} {\bibinfo  {journal} {Nature communications}\ }\textbf {\bibinfo
  {volume} {11}},\ \bibinfo {pages} {4152} (\bibinfo {year}
  {2020})}\BibitemShut {NoStop}%
\bibitem [{\citenamefont {Shen}\ \emph {et~al.}(2017)\citenamefont {Shen},
  \citenamefont {He}, \citenamefont {Yuan}, \citenamefont {Huang},
  \citenamefont {Cho}, \citenamefont {Lee}, \citenamefont {San~Hor},
  \citenamefont {Law},\ and\ \citenamefont {Lortz}}]{shen2017nematic}%
  \BibitemOpen
  \bibfield  {author} {\bibinfo {author} {\bibfnamefont {J.}~\bibnamefont
  {Shen}}, \bibinfo {author} {\bibfnamefont {W.-Y.}\ \bibnamefont {He}},
  \bibinfo {author} {\bibfnamefont {N.~F.~Q.}\ \bibnamefont {Yuan}}, \bibinfo
  {author} {\bibfnamefont {Z.}~\bibnamefont {Huang}}, \bibinfo {author}
  {\bibfnamefont {C.-w.}\ \bibnamefont {Cho}}, \bibinfo {author} {\bibfnamefont
  {S.~H.}\ \bibnamefont {Lee}}, \bibinfo {author} {\bibfnamefont
  {Y.}~\bibnamefont {San~Hor}}, \bibinfo {author} {\bibfnamefont {K.~T.}\
  \bibnamefont {Law}},\ and\ \bibinfo {author} {\bibfnamefont {R.}~\bibnamefont
  {Lortz}},\ }\href@noop {} {\bibfield  {journal} {\bibinfo  {journal} {npj
  Quantum Materials}\ }\textbf {\bibinfo {volume} {2}},\ \bibinfo {pages} {59}
  (\bibinfo {year} {2017})}\BibitemShut {NoStop}%
\bibitem [{\citenamefont {Asaba}\ \emph {et~al.}(2017)\citenamefont {Asaba},
  \citenamefont {Lawson}, \citenamefont {Tinsman}, \citenamefont {Chen},
  \citenamefont {Corbae}, \citenamefont {Li}, \citenamefont {Qiu},
  \citenamefont {Hor}, \citenamefont {Fu},\ and\ \citenamefont
  {Li}}]{asaba2017rotational}%
  \BibitemOpen
  \bibfield  {author} {\bibinfo {author} {\bibfnamefont {T.}~\bibnamefont
  {Asaba}}, \bibinfo {author} {\bibfnamefont {B.}~\bibnamefont {Lawson}},
  \bibinfo {author} {\bibfnamefont {C.}~\bibnamefont {Tinsman}}, \bibinfo
  {author} {\bibfnamefont {L.}~\bibnamefont {Chen}}, \bibinfo {author}
  {\bibfnamefont {P.}~\bibnamefont {Corbae}}, \bibinfo {author} {\bibfnamefont
  {G.}~\bibnamefont {Li}}, \bibinfo {author} {\bibfnamefont {Y.}~\bibnamefont
  {Qiu}}, \bibinfo {author} {\bibfnamefont {Y.~S.}\ \bibnamefont {Hor}},
  \bibinfo {author} {\bibfnamefont {L.}~\bibnamefont {Fu}},\ and\ \bibinfo
  {author} {\bibfnamefont {L.}~\bibnamefont {Li}},\ }\href@noop {} {\bibfield
  {journal} {\bibinfo  {journal} {Physical Review X}\ }\textbf {\bibinfo
  {volume} {7}},\ \bibinfo {pages} {011009} (\bibinfo {year}
  {2017})}\BibitemShut {NoStop}%
\bibitem [{\citenamefont {Le}\ \emph {et~al.}(2020)\citenamefont {Le},
  \citenamefont {Sun}, \citenamefont {Jin}, \citenamefont {Che}, \citenamefont
  {Yin}, \citenamefont {Li}, \citenamefont {Pang}, \citenamefont {Xu},
  \citenamefont {Zhao}, \citenamefont {Kittaka}, \citenamefont {Sakakibara},
  \citenamefont {Machida}, \citenamefont {Sankar}, \citenamefont {Yuan},
  \citenamefont {Chen}, \citenamefont {Xu}, \citenamefont {Li}, \citenamefont
  {Zhou},\ and\ \citenamefont {Lu}}]{XinLu2020evidence}%
  \BibitemOpen
  \bibfield  {author} {\bibinfo {author} {\bibfnamefont {T.}~\bibnamefont
  {Le}}, \bibinfo {author} {\bibfnamefont {Y.}~\bibnamefont {Sun}}, \bibinfo
  {author} {\bibfnamefont {H.-K.}\ \bibnamefont {Jin}}, \bibinfo {author}
  {\bibfnamefont {L.}~\bibnamefont {Che}}, \bibinfo {author} {\bibfnamefont
  {L.}~\bibnamefont {Yin}}, \bibinfo {author} {\bibfnamefont {J.}~\bibnamefont
  {Li}}, \bibinfo {author} {\bibfnamefont {G.}~\bibnamefont {Pang}}, \bibinfo
  {author} {\bibfnamefont {C.}~\bibnamefont {Xu}}, \bibinfo {author}
  {\bibfnamefont {L.}~\bibnamefont {Zhao}}, \bibinfo {author} {\bibfnamefont
  {S.}~\bibnamefont {Kittaka}}, \bibinfo {author} {\bibfnamefont
  {T.}~\bibnamefont {Sakakibara}}, \bibinfo {author} {\bibfnamefont
  {K.}~\bibnamefont {Machida}}, \bibinfo {author} {\bibfnamefont
  {R.}~\bibnamefont {Sankar}}, \bibinfo {author} {\bibfnamefont
  {H.}~\bibnamefont {Yuan}}, \bibinfo {author} {\bibfnamefont {G.}~\bibnamefont
  {Chen}}, \bibinfo {author} {\bibfnamefont {X.}~\bibnamefont {Xu}}, \bibinfo
  {author} {\bibfnamefont {S.}~\bibnamefont {Li}}, \bibinfo {author}
  {\bibfnamefont {Y.}~\bibnamefont {Zhou}},\ and\ \bibinfo {author}
  {\bibfnamefont {X.}~\bibnamefont {Lu}},\ }\href@noop {} {\bibfield  {journal}
  {\bibinfo  {journal} {Science Bulletin}\ }\textbf {\bibinfo {volume} {65}},\
  \bibinfo {pages} {1349} (\bibinfo {year} {2020})}\BibitemShut {NoStop}%
\bibitem [{\citenamefont {Kerelsky}\ \emph {et~al.}(2019)\citenamefont
  {Kerelsky}, \citenamefont {McGilly}, \citenamefont {Kennes}, \citenamefont
  {Xian}, \citenamefont {Yankowitz}, \citenamefont {Chen}, \citenamefont
  {Watanabe}, \citenamefont {Taniguchi}, \citenamefont {Hone}, \citenamefont
  {Dean}, \citenamefont {Rubio},\ and\ \citenamefont
  {Pasupathy}}]{kerelsky2019maximized}%
  \BibitemOpen
  \bibfield  {author} {\bibinfo {author} {\bibfnamefont {A.}~\bibnamefont
  {Kerelsky}}, \bibinfo {author} {\bibfnamefont {L.~J.}\ \bibnamefont
  {McGilly}}, \bibinfo {author} {\bibfnamefont {D.~M.}\ \bibnamefont {Kennes}},
  \bibinfo {author} {\bibfnamefont {L.}~\bibnamefont {Xian}}, \bibinfo {author}
  {\bibfnamefont {M.}~\bibnamefont {Yankowitz}}, \bibinfo {author}
  {\bibfnamefont {S.}~\bibnamefont {Chen}}, \bibinfo {author} {\bibfnamefont
  {K.}~\bibnamefont {Watanabe}}, \bibinfo {author} {\bibfnamefont
  {T.}~\bibnamefont {Taniguchi}}, \bibinfo {author} {\bibfnamefont
  {J.}~\bibnamefont {Hone}}, \bibinfo {author} {\bibfnamefont {C.}~\bibnamefont
  {Dean}}, \bibinfo {author} {\bibfnamefont {A.}~\bibnamefont {Rubio}},\ and\
  \bibinfo {author} {\bibfnamefont {A.~N.}\ \bibnamefont {Pasupathy}},\
  }\href@noop {} {\bibfield  {journal} {\bibinfo  {journal} {Nature}\ }\textbf
  {\bibinfo {volume} {572}},\ \bibinfo {pages} {95} (\bibinfo {year}
  {2019})}\BibitemShut {NoStop}%
\bibitem [{\citenamefont {Cao}\ \emph {et~al.}(2021)\citenamefont {Cao},
  \citenamefont {Rodan-Legrain}, \citenamefont {Park}, \citenamefont {Yuan},
  \citenamefont {Watanabe}, \citenamefont {Taniguchi}, \citenamefont
  {Fernandes}, \citenamefont {Fu},\ and\ \citenamefont
  {Jarillo-Herrero}}]{cao2020nematicity}%
  \BibitemOpen
  \bibfield  {author} {\bibinfo {author} {\bibfnamefont {Y.}~\bibnamefont
  {Cao}}, \bibinfo {author} {\bibfnamefont {D.}~\bibnamefont {Rodan-Legrain}},
  \bibinfo {author} {\bibfnamefont {J.~M.}\ \bibnamefont {Park}}, \bibinfo
  {author} {\bibfnamefont {N.~F.}\ \bibnamefont {Yuan}}, \bibinfo {author}
  {\bibfnamefont {K.}~\bibnamefont {Watanabe}}, \bibinfo {author}
  {\bibfnamefont {T.}~\bibnamefont {Taniguchi}}, \bibinfo {author}
  {\bibfnamefont {R.~M.}\ \bibnamefont {Fernandes}}, \bibinfo {author}
  {\bibfnamefont {L.}~\bibnamefont {Fu}},\ and\ \bibinfo {author}
  {\bibfnamefont {P.}~\bibnamefont {Jarillo-Herrero}},\ }\href@noop {}
  {\bibfield  {journal} {\bibinfo  {journal} {science}\ }\textbf {\bibinfo
  {volume} {372}},\ \bibinfo {pages} {264} (\bibinfo {year}
  {2021})}\BibitemShut {NoStop}%
\bibitem [{\citenamefont {Wu}(1982)}]{wu1982potts}%
  \BibitemOpen
  \bibfield  {author} {\bibinfo {author} {\bibfnamefont {F.-Y.}\ \bibnamefont
  {Wu}},\ }\href@noop {} {\bibfield  {journal} {\bibinfo  {journal} {Rev. Mod.
  Phys.}\ }\textbf {\bibinfo {volume} {54}},\ \bibinfo {pages} {235} (\bibinfo
  {year} {1982})}\BibitemShut {NoStop}%
\bibitem [{\citenamefont {Cho}\ \emph {et~al.}(2020)\citenamefont {Cho},
  \citenamefont {Shen}, \citenamefont {Lyu}, \citenamefont {Atanov},
  \citenamefont {Chen}, \citenamefont {Lee}, \citenamefont {San~Hor},
  \citenamefont {Gawryluk}, \citenamefont {Pomjakushina}, \citenamefont
  {Bartkowiak} \emph {et~al.}}]{cho2020z}%
  \BibitemOpen
  \bibfield  {author} {\bibinfo {author} {\bibfnamefont {C.-w.}\ \bibnamefont
  {Cho}}, \bibinfo {author} {\bibfnamefont {J.}~\bibnamefont {Shen}}, \bibinfo
  {author} {\bibfnamefont {J.}~\bibnamefont {Lyu}}, \bibinfo {author}
  {\bibfnamefont {O.}~\bibnamefont {Atanov}}, \bibinfo {author} {\bibfnamefont
  {Q.}~\bibnamefont {Chen}}, \bibinfo {author} {\bibfnamefont {S.~H.}\
  \bibnamefont {Lee}}, \bibinfo {author} {\bibfnamefont {Y.}~\bibnamefont
  {San~Hor}}, \bibinfo {author} {\bibfnamefont {D.~J.}\ \bibnamefont
  {Gawryluk}}, \bibinfo {author} {\bibfnamefont {E.}~\bibnamefont
  {Pomjakushina}}, \bibinfo {author} {\bibfnamefont {M.}~\bibnamefont
  {Bartkowiak}}, \emph {et~al.},\ }\href@noop {} {\bibfield  {journal}
  {\bibinfo  {journal} {Nature communications}\ }\textbf {\bibinfo {volume}
  {11}},\ \bibinfo {pages} {3065} (\bibinfo {year} {2020})}\BibitemShut
  {NoStop}%
\bibitem [{SM()}]{SM}%
  \BibitemOpen
  \href@noop {} {}\bibinfo {note} {See Supplemental Materials for the technical
  details of the derivation of the saddle-point equation.}\BibitemShut {Stop}%
\bibitem [{\citenamefont {Chichinadze}\ \emph {et~al.}(2020)\citenamefont
  {Chichinadze}, \citenamefont {Classen},\ and\ \citenamefont
  {Chubukov}}]{chichinadze2020nematic}%
  \BibitemOpen
  \bibfield  {author} {\bibinfo {author} {\bibfnamefont {D.~V.}\ \bibnamefont
  {Chichinadze}}, \bibinfo {author} {\bibfnamefont {L.}~\bibnamefont
  {Classen}},\ and\ \bibinfo {author} {\bibfnamefont {A.~V.}\ \bibnamefont
  {Chubukov}},\ }\href@noop {} {\bibfield  {journal} {\bibinfo  {journal}
  {Physical Review B}\ }\textbf {\bibinfo {volume} {101}},\ \bibinfo {pages}
  {224513} (\bibinfo {year} {2020})}\BibitemShut {NoStop}%
\bibitem [{\citenamefont {L{\"o}thman}\ \emph {et~al.}(2021)\citenamefont
  {L{\"o}thman}, \citenamefont {Schmidt}, \citenamefont {Parhizgar},\ and\
  \citenamefont {Black-Schaffer}}]{lothman2021nematic}%
  \BibitemOpen
  \bibfield  {author} {\bibinfo {author} {\bibfnamefont {T.}~\bibnamefont
  {L{\"o}thman}}, \bibinfo {author} {\bibfnamefont {J.}~\bibnamefont
  {Schmidt}}, \bibinfo {author} {\bibfnamefont {F.}~\bibnamefont {Parhizgar}},\
  and\ \bibinfo {author} {\bibfnamefont {A.~M.}\ \bibnamefont
  {Black-Schaffer}},\ }\href@noop {} {\bibfield  {journal} {\bibinfo  {journal}
  {arXiv preprint arXiv:2101.11555}\ } (\bibinfo {year} {2021})}\BibitemShut
  {NoStop}%
\bibitem [{\citenamefont {Sherkunov}\ and\ \citenamefont
  {Betouras}(2018)}]{sherkunov2018electronic}%
  \BibitemOpen
  \bibfield  {author} {\bibinfo {author} {\bibfnamefont {Y.}~\bibnamefont
  {Sherkunov}}\ and\ \bibinfo {author} {\bibfnamefont {J.~J.}\ \bibnamefont
  {Betouras}},\ }\href@noop {} {\bibfield  {journal} {\bibinfo  {journal}
  {Physical Review B}\ }\textbf {\bibinfo {volume} {98}},\ \bibinfo {pages}
  {205151} (\bibinfo {year} {2018})}\BibitemShut {NoStop}%
\bibitem [{\citenamefont {Gonzalez}\ and\ \citenamefont
  {Stauber}(2019)}]{gonzalez2019kohn}%
  \BibitemOpen
  \bibfield  {author} {\bibinfo {author} {\bibfnamefont {J.}~\bibnamefont
  {Gonzalez}}\ and\ \bibinfo {author} {\bibfnamefont {T.}~\bibnamefont
  {Stauber}},\ }\href@noop {} {\bibfield  {journal} {\bibinfo  {journal}
  {Physical review letters}\ }\textbf {\bibinfo {volume} {122}},\ \bibinfo
  {pages} {026801} (\bibinfo {year} {2019})}\BibitemShut {NoStop}%
\bibitem [{\citenamefont {Lian}\ \emph {et~al.}(2019)\citenamefont {Lian},
  \citenamefont {Wang},\ and\ \citenamefont {Bernevig}}]{lian2019twisted}%
  \BibitemOpen
  \bibfield  {author} {\bibinfo {author} {\bibfnamefont {B.}~\bibnamefont
  {Lian}}, \bibinfo {author} {\bibfnamefont {Z.}~\bibnamefont {Wang}},\ and\
  \bibinfo {author} {\bibfnamefont {B.~A.}\ \bibnamefont {Bernevig}},\
  }\href@noop {} {\bibfield  {journal} {\bibinfo  {journal} {Physical review
  letters}\ }\textbf {\bibinfo {volume} {122}},\ \bibinfo {pages} {257002}
  (\bibinfo {year} {2019})}\BibitemShut {NoStop}%
\bibitem [{\citenamefont {Fischer}\ \emph {et~al.}(2021)\citenamefont
  {Fischer}, \citenamefont {Klebl}, \citenamefont {Honerkamp},\ and\
  \citenamefont {Kennes}}]{fischer2021spin}%
  \BibitemOpen
  \bibfield  {author} {\bibinfo {author} {\bibfnamefont {A.}~\bibnamefont
  {Fischer}}, \bibinfo {author} {\bibfnamefont {L.}~\bibnamefont {Klebl}},
  \bibinfo {author} {\bibfnamefont {C.}~\bibnamefont {Honerkamp}},\ and\
  \bibinfo {author} {\bibfnamefont {D.~M.}\ \bibnamefont {Kennes}},\
  }\href@noop {} {\bibfield  {journal} {\bibinfo  {journal} {Physical Review
  B}\ }\textbf {\bibinfo {volume} {103}},\ \bibinfo {pages} {L041103} (\bibinfo
  {year} {2021})}\BibitemShut {NoStop}%
\bibitem [{\citenamefont {Fernandes}\ and\ \citenamefont
  {Fu}(2021)}]{fernandes2021charge}%
  \BibitemOpen
  \bibfield  {author} {\bibinfo {author} {\bibfnamefont {R.~M.}\ \bibnamefont
  {Fernandes}}\ and\ \bibinfo {author} {\bibfnamefont {L.}~\bibnamefont {Fu}},\
  }\href {https://doi.org/10.1103/PhysRevLett.127.047001} {\bibfield  {journal}
  {\bibinfo  {journal} {Phys. Rev. Lett.}\ }\textbf {\bibinfo {volume} {127}},\
  \bibinfo {pages} {047001} (\bibinfo {year} {2021})}\BibitemShut {NoStop}%
\end{thebibliography}%

\onecolumngrid
\vspace{1cm}
\begin{center}
{\bf\large Supplemental Material}
\end{center}

\setcounter{secnumdepth}{3}
\setcounter{equation}{0}
\setcounter{figure}{0}
\renewcommand{\theequation}{S\arabic{equation}}
\renewcommand{\thefigure}{S\arabic{figure}}
\renewcommand\figurename{Supplementary Figure}
\renewcommand\tablename{Supplementary Table}
\newcommand\Scite[1]{[S\citealp{#1}]}
\makeatletter \renewcommand\@biblabel[1]{[S#1]} \makeatother

There are two ways to decouple the quartic term in Eq.~(\ref{eq:nematic_SC}), given by the following two different Fierz identities 
\bea
    \tau^y_{\alpha\beta} \tau^y_{\gamma\delta} &=& 2 \delta_{\alpha\delta} \delta_{\beta\gamma} - \delta_{\alpha\beta} \delta_{\gamma\delta}- \tau^x_{\alpha\beta} \tau^x_{\gamma\delta} - \tau^z_{\alpha\beta} \tau^z_{\gamma\delta}, \\
    \tau^y_{\alpha\beta} \tau^y_{\gamma\delta} &=& \delta_{\alpha\beta} \delta_{\gamma\delta} - \delta_{\alpha \gamma} \delta_{\beta\delta},
\eea
which lead to two exact rewritings of Eq.~(\ref{eq:nematic_SC})
\bea
\label{eq:nematic_H}	S &=&  \int_p \Delta^\dag G^{-1}_p \Delta + \int_x( u' (\Delta^\dag \Delta)^2 - v [(\Delta^\dag \tau^x \Delta)^2 + (\Delta^\dag \tau^z \Delta)^2]), \\
\label{eq:4e_H}	S &=&  \int_p \Delta^\dag G^{-1}_p \Delta + \int_x( u' (\Delta^\dag \Delta)^2 - v |\Delta^T  \Delta|^2),
\eea
where $u'=u+v$.

The first equation~(\ref{eq:nematic_H}) is readily decoupled by nematic orders, $(Q_x, Q_y)$,
\bea
    S &=& \int_x \left( \Delta^\dag (G^{-1} + R + Q_x \tau^z + Q_y \tau^x) \Delta  + \frac1{4v} (Q_x^2 + Q_y^2) - \frac1{4u'} R^2) \right),
\eea
where $R$ is another boson field that decouple $u'$ term, and it does not break any symmetry.
On the other hand, the second equation~(\ref{eq:4e_H}) is readily decoupled by charge-$4e$ SC orders, $(\Delta_{4e}, \Delta_{4e}^\ast)$,
\bea
    S &=& \int_x  \left( \frac12 \hat\Delta^\dag \left( \ba{cccc} G^{-1} + R &  \Delta_{4e} \\  \Delta_{4e}^\ast  &  G^{-1} + R \ea \right) \hat \Delta  + \frac1{4v} |\Delta_{4e}|^2 - \frac1{4u'} R^2 \right)
\eea
where $\hat \Delta = (\Delta^\dag, \Delta^T) $.

Since now the nematic SC order is quadratic in the action, it is straightforward to integrate it over. 
\bea
    S &=& \Tr\log (G^{-1} + R + Q_x \tau^z + Q_y \tau^x) + \int_x \left( \frac1{4v} (Q_x^2 + Q_y^2) - \frac1{4u'} R^2 \right), \\
    S &=& \frac12 \Tr \log \left( \ba{cccc} G^{-1} + R &  \Delta_{4e} \\  \Delta_{4e}^\ast  &  G^{-1} + R \ea \right) + \int_x \left( \frac1{4v} |\Delta_{4e}|^2 - \frac1{4u'} R^2 \right).
\eea

From the effective action~(\ref{eq:nematic_action}), the saddle-point equation for homogeneous nematic orders, $Q_i(\bm x) =  Q_i$, $i=x,y$ reads
\bea
	R &=& 2u' \Delta_0^\dag \Delta_0 + 2u' \int_p \Tr (G_p^{-1} + R + Q_x \tau^z + Q_y \tau^x)^{-1},  \\
	Q_x &=& -2v \Delta_0^\dag \tau^z \Delta_0 -2v \int_p \Tr (G_p^{-1} + R + Q_x \tau^z + Q_y \tau^x)^{-1} \tau^z, \\
	Q_y &=& -2v \Delta_0^\dag \tau^x \Delta_0 -2v \int_p \Tr (G_p^{-1} + R + Q_x \tau^z + Q_y \tau^x)^{-1} \tau^x, \\
	0 &=& \chi_0^{-1} \Delta_0.
\eea
On the other hand, the saddle-point equation for homogeneous charge-$4e$ SC orders $\Delta_{4e}(\bm x) =  \Delta_{4e}'+ i \Delta_{4e}''$ where $\Delta_{4e}' , \Delta_{4e}'' \in \mathbb{R}$ from Eq.~(\ref{eq:4e_action}) reads
\bea
    R &=& 2u' \Delta_0^\dag \Delta_0 +  u' \int_p \Tr \left( \ba{cccc} G^{-1}_p + R &  \Delta_{4e}'+ i \Delta_{4e}'' \\  \Delta_{4e}'- i \Delta_{4e}''  &  G^{-1}_p + R \ea \right)^{-1}  \\
	\Delta_{4e}' &=&  -2v \text{Re}[\Delta_0^T \Delta_0] -  v \int_p \Tr \left( \ba{cccc} G^{-1}_p + R &  \Delta_{4e}'+ i \Delta_{4e}'' \\  \Delta_{4e}'- i \Delta_{4e}''  &  G^{-1}_p + R \ea \right)^{-1} \rho^x, \\
	\Delta_{4e}'' &=& -2v \text{Im}[\Delta_0^T \Delta_0] -  v \int_p \Tr \left( \ba{cccc} G^{-1}_p + R &  \Delta_{4e}'+ i \Delta_{4e}'' \\  \Delta_{4e}'- i \Delta_{4e}''  &  G^{-1}_p + R \ea \right)^{-1} (-\rho^y), \\
	0 &=& \chi_0^{-1} \Delta_0.
\eea
In the saddle-point equations, we also assume $R(\bm x) = R$ is homogeneous. 
The saddle-point equations are solved numerically, and the results are shown in Fig.~\ref{fig:meanfield}.

\end{document}